\documentclass[prl,superscriptaddress,twocolumn]{revtex4}
\usepackage{graphicx}

\def\lno{La$_2$NiO$_4$}
\def\lsno{La$_{2-x}$Sr$_x$NiO$_4$}
\def\lsco{La$_{2-x}$Sr$_x$CuO$_4$}

\def\vq{{\bf Q}}
\def\vqo{{\bf Q}_0}
\def\tm{T_{\rm so}}
\def\tco{T_{\rm co}}

\begin{document}

\title{Freezing of a Stripe Liquid}
\author{S.-H. Lee}
\affiliation{NIST Center for Neutron Research, National Institute of
Standards and Technology, Gaithersburg, MD 20899}
\affiliation{University of Maryland, College Park, MD 20742}
\author{J.M.~Tranquada}
\affiliation{Brookhaven National Laboratory, Upton, NY 11973}
\author{K. Yamada}
\affiliation{Institute for Chemical Research, Kyoto University, Gokashou,
Uji, 611-0011 Kyoto, Japan}
\author{D.J.~Buttrey}
\affiliation{Department of Chemical Engineering, University of Delaware,
Newark, Delaware 19716}
\author{Q. Li}
\affiliation{Brookhaven National Laboratory, Upton, NY 11973}
\author{S.-W. Cheong}
\affiliation{Department of Physics \&\ Astronomy, Rutgers University,
Piscataway, NJ 08855-0849}
\affiliation{Bell Laboratories, Lucent Technologies, Murray Hill, NJ
07974}
\date{\today}
\begin{abstract}
The existence of a stripe-liquid phase in a layered nickelate,
La$_{1.725}$Sr$_{0.275}$NiO$_4$, is demonstrated through neutron
scattering measurements.  We show that incommensurate magnetic
fluctuations evolve continuously through the charge-ordering temperature,
although an abrupt decrease in the effective damping energy is observed on
cooling through the transition.  The energy and momentum dependence of
the magnetic scattering are parametrized with a damped-harmonic-oscillator
model describing overdamped spin-waves in the antiferromagnetic domains
defined instantaneously by charge stripes.
\end{abstract}
\pacs{PACS: 71.27.+a, 75.40.Cx, 75.50.Ee, 71.45.Lr}
\maketitle

One of the key issues in current debates over copper-oxide
superconductors concerns the nature and relevance of charge stripes
\cite{emer99}.  It has been demonstrated in one cuprate family that the
holes doped into the CuO$_2$ planes can order in an array of
periodically-spaced stripes, separating antiferromagnetic domains
\cite{ichi00}.  For charge stripes to be relevant to superconductivity,
they must be ubiquitous among the cuprates, and for the latter to be true,
they must be able to exist in a liquid state.  Indeed, theoretical models
for such electronic liquid crystal phases have been proposed
\cite{zaan96a,kive98}.  The inelastic incommensurate magnetic scattering
observed in La$_{2-x}$Sr$_x$CuO$_4$ \cite{cheo91,yama98a} and
YBa$_2$Cu$_3$O$_{6+x}$ \cite{mook98} has sometimes been interpreted as
evidence for dynamic stripes; however, this interpretation has been
controversial \cite{bour00}.

Here we present evidence for a stripe liquid phase in a related system,
La$_{2-x}$Sr$_x$NiO$_4$.  Ordered stripes have been observed in this
system over a large range of hole concentrations \cite{yosh00}.  Although
the maximum charge and spin ordering temperatures occur for $x=0.33$
\cite{lee97}, we have chosen to study crystals of $x=0.275$, which have
the advantage that the spin and charge-ordering wave vectors do not
coincide.  Using neutron scattering, we follow the magnetic inelastic
scattering to temperatures as high as 1.7 times the charge-ordering
transition, $\tco$, and demonstrate the existence of a stripe liquid. 
(\lsno\ is nonmetallic even at $T\gg\tco$, as indicated by a peak in its
optical conductivity at
$\sim0.5$~eV \cite{kats96,pash00}, so there can be little controversy over
possible alternative interpretations involving Fermi-surface nesting.) 
We show that the
$\omega$ and $\vq$ dependence of the data can be effectively parametrized
with a damped-harmonic-oscillator (DHO) model describing overdamped spin
waves associated with the antiferromagnetic domains defined
instantaneously by the charge stripes.  Using a spin-wave energy
dispersion with an effective gap energy, we find that the gap energy is
proportional to the inverse correlation length and varies linearly
with temperature through $\tco$.  On the other hand, the behavior
of the damping parameter changes abruptly near $\tco$.

The neutron scattering measurements were performed on the SPINS
triple-axis spectrometer in the cold-neutron guide hall at the NIST
Center for Neutron Research (NCNR).  Initially, a single crystal of 4.4~g
was used; later, a second crystal of 6.3~g was mounted along side the
first.  Both crystals were grown at Kyoto University, and annealed at the
University of Delaware in order to achieve a stoichiometric oxygen
concentration. (The first crystal was used previously for the work
reported in \cite{lee01}, but without annealing.)  The neutron
spectrometer was equipped with a vertically-focusing monochromator and
horizontally-focusing analyzer, both utilizing PG (002) reflections. 
Elastic scans of superlattice peaks were performed with 5-meV neutrons. 
Inelastic measurements were done in energy-gain mode, with incident
energies of either 5 or 13.7 meV.  The sample temperature was
controlled with a displex refrigerator.

The stripe order in this sample is consistent with that observed in
previous studies of \lsno\ \cite{yosh00}.  Figure 1(a) shows the
temperature dependence of the spin and charge-order superlattice peaks. 
The spin order approaches zero at $\tm\sim120$~K, while the charge
order disappears by $\tco\sim190$~K.

\begin{figure}[t]
\centerline{\includegraphics[width=3.2in]{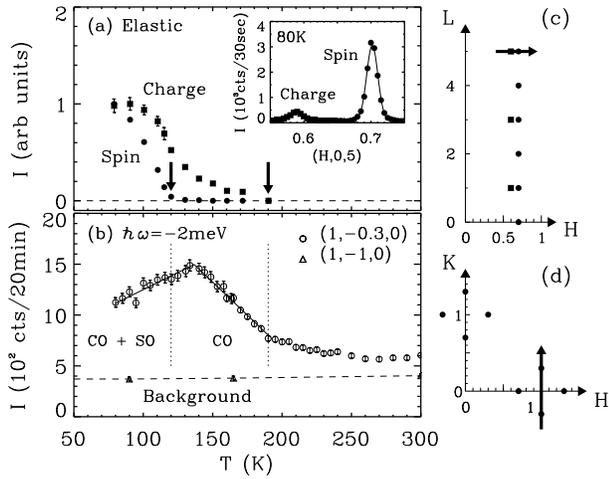}}
\medskip
\caption{(a) Temperature dependence of spin-order (SO) and charge-order
(CO) superlattice intensities.  Arrows indicate approximate transition
temperatures.  Inset: scan of intensity vs.\ {\bf Q} through superlattice
peaks at $(2\epsilon,0,5)$ and $(1-\epsilon,0,5)$ showing that
$\epsilon\approx0.29$ at 80 K.  (b) Intensity at $(1,-\epsilon,0)$ for an
energy gain of 2 meV as a function of temperature.  Points
sampled at $(1,-1,0)$ indicate the background.  The lines through points
are guides to the eye. (c) $(H0L)$ zone of reciprocal space, showing
positions of magnetic (circles) and charge-order (squares) peaks.  Arrow
indicates direction of scan in inset of (a).  (d) $(HK0)$ zone, showing
magnetic peaks.  Arrow indicates scan direction corresponding to
Fig.~2(a)--(c).}
\label{fg:1}
\end{figure}

Our focus here is on the dynamic correlations, especially those at
$T>\tco$.  Figure 2 shows examples of constant-energy-transfer
scans ($\hbar\omega = -4$~meV) through the incommensurate magnetic peak
positions $(1,\pm\epsilon,0)$ at temperatures (a) $T>\tco$,
(b) $\tco>T>\tm$, and (c) $\tm>T$.  We clearly observe well-resolved
peaks at all temperatures.  Given that the ordered state at $T<\tco$ is
already well characterized in terms of periodically spaced charge
stripes separating antiferromagnetic domains, the continuous evolution of
the inelastic magnetic scattering [see Fig.~1(b)] provides direct evidence
for the  existence of instantaneously correlated magnetic domains in the
disordered state.  The fact that the scattering at the commensurate
antiferromagnetic wave vector (1,0,0) is always a local minimum indicates
that neighboring antiferromagnetic domains maintain their antiphase
relationship, which is caused by the segregation of the doped holes to
the domain walls.  Thus, we feel that Fig.~2(a) is firm evidence for a
stripe-liquid phase.

\begin{figure}[t]
\centerline{\includegraphics[width=3.2in]{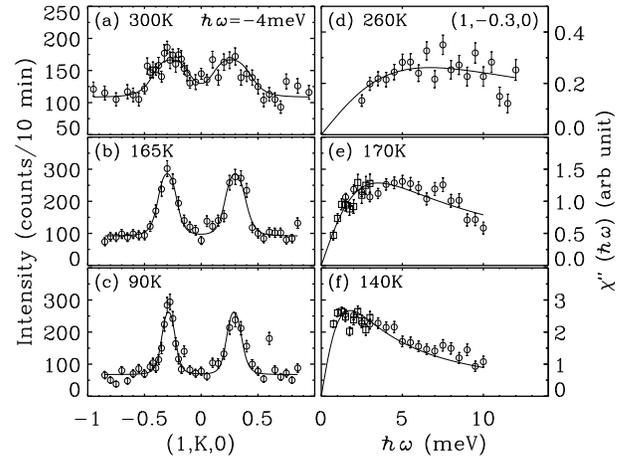}}
\medskip
\caption{Left side: constant-energy scans for $\hbar\omega=-4$~meV for
three different conditions: (a) 300~K, no static order; (b) 165~K, charge
but no spin order; (c) 90~K, spin and charge order.  Right side:
constant-$\vq$ scans at $\vq=(1,-0.3,0)$ for temperatures (d) 260~K, (e)
170~K, (f) 140~K.  The curves through the data are fits as described in
the text.}
\label{fg:2}
\end{figure}

The variation of the $Q$-widths of the peaks in the constant-$E$ scans for
$\hbar\omega=0$, $-4$, and $-8$ meV is plotted in Fig.~3.  For
$T\lesssim\tm$, the widths are roughly temperature independent but vary
substantially with energy.  The variation with energy is similar to what
one might expect from spin-wave dispersion, as observed previously
\cite{tran97c}.  At higher temperatures, the widths appear to vary
linearly with temperature, and the dependence on frequency is reduced.

\begin{figure}[t]
\centerline{\includegraphics[width=3.2in]{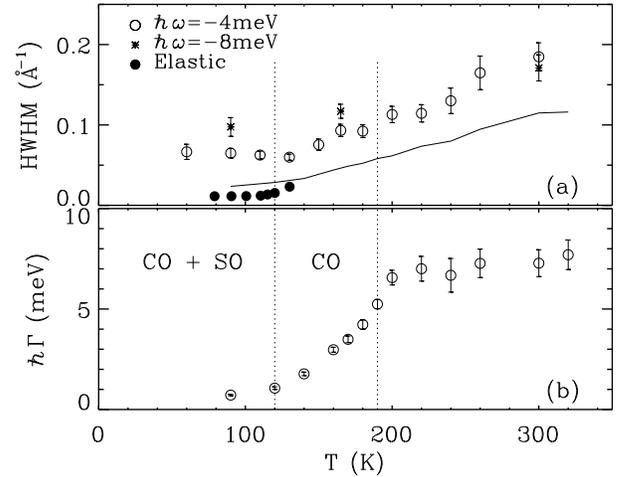}}
\medskip
\caption{(a) Half-width at half maximum (HWHM), without resolution
correction, for constant-$E$ scans through magnetic peaks:
$\hbar\omega=-8$~meV (stars);
$-4$~meV (open circles); elastic (filled circles).  Solid line
corresponds to fit parameter $\omega_0/c$ ($\approx\kappa$). (b) Results
for effective spin-wave damping, $\Gamma$, from fits of DHO model (see
text).}
\label{fg:3}
\end{figure}

The distribution of the magnetic-scattering strength with frequency
also evolves with temperature.  Figure~2(d)-(f) show the variation of the
imaginary part of the dynamic susceptibility, $\chi''(\vq,\omega)$, at
several temperatures; here, ${\bf Q}={\bf Q}_0$, where
${\bf Q}_0$ stands for an incommensurate magnetic wave vector.  $\chi''$
is related to the experimentally measured $S(\vq,\omega)$ via
\begin{equation}
  S(\vq,\omega) = \left(1-e^{-\hbar\omega/k_{\rm B}T}\right)^{-1}
  \chi''(\vq,\omega);
\end{equation}
the correction for the detailed-balance factor was applied after
subtracting the background, measured at
$\vq=(1,-0.9,0)$.  The position of the maximum of
$\chi''$ as a function of energy corresponds to a characteristic damping
energy $\Gamma$, and one can see that $\Gamma$ increases with temperature.

To describe the data more quantitatively, we have chosen to use the
damped-harmonic-oscillator (DHO) model:
\begin{equation}
  \chi''(\vq,\omega) = \sum_{\vqo}{2\omega\gamma\chi_0 \over
    (\omega^2-\omega_\vq^2)^2 + (2\omega\gamma)^2},
    \label{eq:DHO}
\end{equation}
where the excitations are assumed to have the dispersive form
\begin{equation}
  \omega_\vq^2 = \omega_0^2 + c^2(\vq-\vqo)^2.
\end{equation}
(Direct evidence for dispersive modes at high energies
will be presented in a complementary study \cite{bour01}.)  To fit the
data, the model
$S(\vq,\omega)$ was convolved with the spectrometer resolution
function.   The fitting parameters are $\omega_0$, $\gamma$, and
$\chi_0$, with the effective spin-wave velocity held fixed at $\hbar
c=300$~meV~\AA, the approximate value determined for pure \lno\
\cite{yama91}.  To describe
the $\vq$ dependence of the data, the relevant combinations are
$\omega_0/c$ and $\gamma/c$; for the energy dependence, as we will
discuss, it is only the ratio $\omega_0^2/2\gamma$ that matters. 

The solid curves in Fig.~2 represent the fits to the data.  The
parameter values obtained from such fits are displayed in Fig.~4.  For
$T>\tm$, $\omega_0$ varies linearly with temperature.  In contrast,
$\gamma$ is roughly constant up to $T\sim\tco$ where it begins to
increase rapidly.  $\chi_0$ has a fairly weak temperature dependence
which is described approximately by $A/(1+\gamma/B)$, where
$B\approx120$~meV.  The temperature dependence of $\gamma$ suggests that
a major contribution to this damping factor comes from the fluctuations
of the charge stripes \cite{zaan97a}; it is consistent with the growth in
the inverse correlation length for charge stripes observed by x-ray
scattering in
\lsno\ with $x=\frac13$ \cite{du00}.  The saturation of $\gamma$ below
$\tco$ might be due in part to the quenched disorder associated with
the randomly-distributed Sr dopants.  

\begin{figure}[t]
\centerline{\includegraphics[width=3.2in]{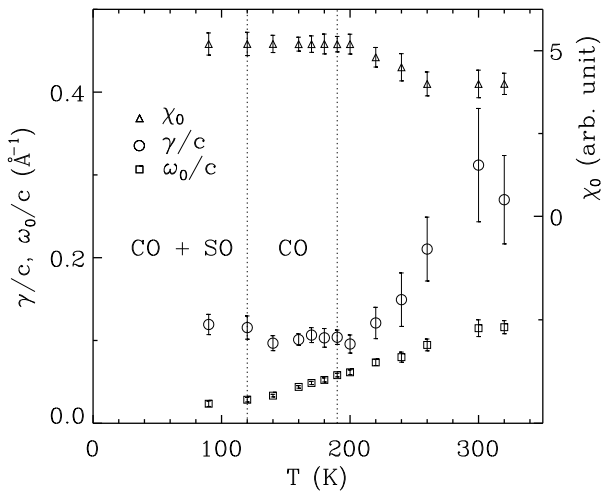}}
\medskip
\caption{Results for fitting parameters vs.\ temperature: $\chi_0$
(triangles, right-hand scale), $\gamma/c$ (circles), and $\omega_0/c$
(squares).}
\label{fg:4}
\end{figure}

Our scans versus frequency (Fig.~2) are measured at $\vq=\vqo$, and most
of the data correspond to $\omega^2\ll\omega_0^2$, in which case
Eq.~(\ref{eq:DHO}) simplifies to
\begin{equation}
  \chi''(\vqo,\omega) \approx {\chi_0\over2\gamma}\cdot
    {\omega\over\omega^2+\Gamma^2},
\end{equation}
where $\Gamma = \omega_0^2/2\gamma$.  The values of $\Gamma$ obtained
from the values of the fitted parameters are plotted in Fig.~3(b).  At
$T>\tco$, $\Gamma$ changes slowly and has a value of roughly 7~meV.  On
cooling through $\tco$, $\Gamma$ abruptly starts to decrease, and
becomes rather small by $\tm$.  Since $\omega_0$ varies smoothly through
this regime, the change in behavior at $\sim\tco$ is controlled by
$\gamma$.  If $\gamma$ is a measure of charge-stripe fluctuations, as
discussed above, then it appears that the abrupt decrease in $\Gamma$
below $\tco$ may be a direct result of charge order.  Low-frequency
probes such as nuclear magnetic resonance \cite{abu99,yosh99} and
muon-spin relaxation \cite{chow96,jest99} should be sensitive to the
variations in $\Gamma$, and it appears that such variations above
$\tm$ have been detected in studies of related nickelates
\cite{abu99,chow96,jest99}.  The sensitivity of magnetic damping to
charge order may also be relevant to the mechanism of the ``wipe-out''
effect observed in nuclear-quadrupole-resonance studies in cuprates
\cite{hunt01,curr00,juli01}.

In order to get further insight into the significance of the model
parameters, we numerically integrated $S(\vq,\omega)$ over frequency. 
The result corresponds to $\vq$-dependent peaks with a line shape that is
approximately Lorentzian.  The half-width-at-half-maximum of this
function should correspond to the instantaneous inverse correlation
length, $\xi^{-1}\equiv\kappa$, and we find that, to within 1\%,
$\kappa=0.86\omega_0/c$.  This result, that the effective spin-excitation
gap $\omega_0$ is approximately equal to $\kappa c$, is equivalent to the 
form proposed for the paramagnetic state of the undoped Heisenberg
antiferromagnet \cite{auer88,mata89}.

It is of interest to consider the transport properties of \lsno\ in the
regime $T>\tco$ which we now associate with the stripe-liquid phase;
numerous single-crystal studies of transport and optical properties have
been reported previously (see, e.g., \cite{kats96,pash00,kats99}). 
Figure~5 compares the temperature-dependence of the in-plane resistivity
measured on our crystal with similar measurements on two \lsco\
samples \cite{ichi00,waki01}.  We observe that the resistivity of our
nickelate crystal is only an order of magnitude greater than that of
underdoped cuprates at 300~K, and we expect that it will develop a
``metallic'' temperature derivative at higher temperatures, as observed
above 400~K in \lsno\ with $x=0.33$ \cite{kats96}.  Furthermore, at high
temperatures the resistivities of all three samples exceed a critical
limit; to the extent that $d\rho/dT$ is positive, they qualify as ``bad''
metals \cite{emer95b}.  We believe it is plausible that the ``bad''
metallic behavior of the cuprates might be associated with a
stripe-liquid phase.  (Resistivity in stripe-ordered cuprates can be
low \cite{ichi00}.)

\begin{figure}[t]
\centerline{\includegraphics[width=3.2in]{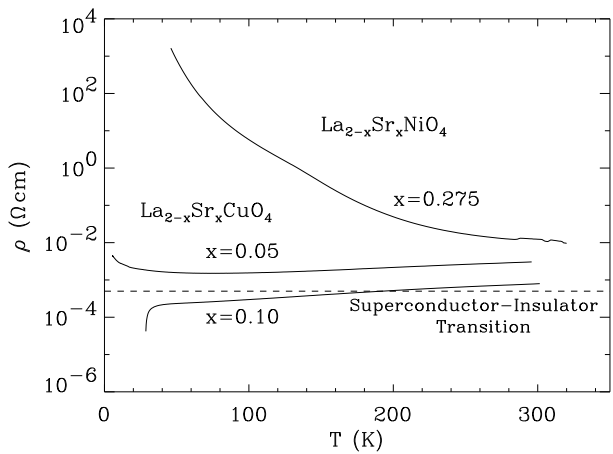}}
\medskip
\caption{Resistivity of the present sample compared with that of
La$_{2-x}$Sr$_x$CuO$_4$ with $x=0.05$ \cite{waki01} and $x=0.10$ 
\protect\cite{ichi00}.  Dashed line indicates the empirically-determined
critical resistivity for the superconductor-insulator transition in layered
cuprates \protect\cite{fuku96,semb01}.  (If the in-plane resistivity does
not drop below the critical range of $\rho\sim0.4$--0.8~m$\Omega$~cm, then
the sample will not go superconducting at any temperature.) }
\label{fg:5}
\end{figure}

The comparable magnitudes of room-temperature resistivity in the
nickelates and cuprates make it plausible that a stripe-liquid phase
could be relevant to transport properties in both materials.  There are
certainly differences in the two systems: the stripes inevitably order in
the nickelates, whereas stripe order is generally avoided in the
cuprates.  However, these differences might be associated more with the
magnitude of fluctuations rather than the nature of instantaneous
correlations.  In \lsco\ with $x=0.14$, the measurements of Aeppli and
coworkers \cite{aepp97} indicate that, using the present parametrization,
$\hbar\Gamma\approx10$~meV at $T=35$~K, with $\Gamma$ increasing
substantially at higher temperatures.  In that case, the relevance of
quantum critical phenonema has been proposed \cite{cast97,sach00}.

To conclude, we have presented experimental evidence for the existence of
a liquid phase of charge stripes in a hole-doped nickelate.  The
magnetic correlations evolve smoothly through the freezing transition,
although the damping of the fluctuations does not.  The in-plane
resistivity, though non-metallic, is relatively low in the stripe-liquid
phase. 

%
%
We gratefully acknowledge helpful comments from S. A. Kivelson and J.
Zaanen.  KY acknowledges support from the Japan Science and Technology
Corporation, the Core Research for Evolutional Science and Technology
Project (CREST), and Grants-in-Aid for Scientific Research on Priority
Areas, 12046239, 2001 and Research (A), 10304026, 2001 and for Creative
Scientific Research (13NP0201) from the Japanese Ministry of Education,
Culture, Sports, Science, and Technology.  Work at SPINS is based upon
activities supported by the National Science Foundation under Agreement
No.\ DMR-9986442. Research at Brookhaven is  supported by the Department
of Energy's (DOE) Office of Science under Contract No.\
DE-AC02-98CH10886.  DJB acknowledges support from DOE under Contract No.\
DE-FG02-00ER45800.


\end{document}